\documentclass[fleqn,usenatbib]{mnras}
\usepackage[utf8]{inputenc}
\usepackage{newtxtext,newtxmath}

\usepackage[T1]{fontenc}
\usepackage{ae,aecompl}
\usepackage[mathscr]{euscript}
\usepackage{graphicx}
\usepackage{subfig}
\usepackage{amsmath}
\usepackage{amssymb}
\graphicspath{{./Images/}}
\usepackage{deluxetable}
\usepackage{ulem}
\newcommand{\msun}{$\rm M_{\odot}$}
\newcommand{\rsun}{$\rm R_{\odot}$}
\newcommand{\beq}{\begin{equation}}
\newcommand{\eeq}{\end{equation}}
\newcommand{\beqn}{\begin{eqnarray}}
\newcommand{\eeqn}{\end{eqnarray}}

\makeatletter
\def\@to{to}
\makeatother

\begin{document}\label{firstpage}\setcounter{page}{5393}
\volume{522}\pagerange{5393-5401}\pubyear{2023}
\title[Aligning Retrograde NSC Orbits in AGN]{Aligning Retrograde Nuclear Cluster Orbits with an Active Galactic Nucleus Accretion Disc}
\author[S. S. Nasim et al.]{Syeda S. Nasim,$^{1,2}$\thanks{E-mail: ssnasim@mst.edu (SSN); gaia.fabj@nbi.ku.dk (GF)} Gaia Fabj,$^{2,3\star}$ Freddy Caban,$^{2,4}$ Amy Secunda,$^{2,5}$
\newauthor K. E. Saavik Ford,$^{2,6,7}$ Barry McKernan,$^{2,6,7}$ Jillian M. Bellovary,$^{2,8,9}$
\newauthor Nathan W. C. Leigh,$^{2,10}$ and Wladimir Lyra$^{11}$\\
$^{1}$Department of Physics, Missouri University of Science and Technology, Rolla, MO 65409, USA\\
$^{2}$Department of Astrophysics, American Museum of Natural History, New York, NY 10024, USA\\
$^{3}$Niels Bohr International Academy, The Niels Bohr Institute, Blegdamsvej 17, 2100 Copenhagen, Denmark \\
$^{4}$Department of Physics, Stony Brook University, State University of New York, Stony Brook, NY 11794, USA\\
$^{5}$Department of Astrophysical Sciences, Princeton University, Peyton Hall, Princeton, NJ 08544, USA\\
$^{6}$Department of Science, Borough of Manhattan Community College, City University of New York, New York, NY 10007, USA\\
$^{7}$Physics Program, CUNY Graduate Center, City University of New York, New York, NY 10016, USA\\
$^{8}$Astrophysics Program, CUNY Graduate Center, City University of New York, New York, NY 10016, USA\\
$^{9}$Department of Physics, Queensborough Community College, City University of New York, Bayside, NY 11364, USA\\
$^{10}$Departamento de Astronom\'ia, Facultad Ciencias F\'isicas y Matem\'aticas, Universidad de Concepci\'on, 403000 Concepci\'on, Chile\\
$^{11}$Department of Astronomy, New Mexico State University, Las Cruces, NM 88003, USA}
\date{Accepted 2023 March 27. Received 2023 March 13; in original form 2022 July 18. ~\href{https://doi.org/10.1093/mnras/stad1295}{doi:10.1093/mnras/stad1295}}

\maketitle

\begin{abstract} Stars and stellar remnants orbiting a supermassive black hole (SMBH) can interact with an active galactic nucleus (AGN) disc. Over time, prograde orbiters (inclination $i<90^{\circ}$) decrease inclination, as well as semi-major axis $(a)$ and eccentricity $(e)$ until orbital alignment with the gas disc (‘disc capture’). Captured stellar-origin black holes (sBH) add to the embedded AGN population which drives sBH-sBH mergers detectable in gravitational waves using LIGO-Virgo-KAGRA (LVK) or sBH-SMBH mergers detectable with LISA (Laser Interferometer Space Antenna). Captured stars can be tidally disrupted by sBH or the SMBH or rapidly grow into massive ‘immortal’ stars. Here, we investigate the behaviour of polar and retrograde orbiters $(i \geq 90^{\circ})$ interacting with the disc. We show that retrograde stars are captured faster than prograde stars, flip to prograde orientation $(i<90^{\circ})$ during capture, and decrease $a$ dramatically towards the SMBH. For sBH, we find a critical angle $i_{\rm ret} \sim 113^{\circ}$, below which retrograde sBH decay towards embedded prograde orbits $(i \to 0^{\circ})$, while for $i_{\rm o}>i_{\rm ret}$ sBH decay towards embedded retrograde orbits $(i \to 180^{\circ})$. sBH near polar orbits $(i \sim 90^{\circ})$ and stars on nearly embedded retrograde orbits $(i \sim 180^{\circ})$ show the greatest decreases in $a$. Whether a star is captured by the disc within an AGN lifetime depends primarily on disc density, and secondarily on stellar type and initial $a$. For sBH, disc capture-time is longest for polar orbits, low mass sBH and lower density discs. Larger mass sBH should typically spend more time in AGN discs, with implications for the embedded sBH spin distribution. \end{abstract}


\section{Introduction}\label{sec:Introduction}
Active galactic nuclei (AGN) are believed to be powered by supermassive black holes (SMBH) accreting from luminous gas discs. Before the AGN disc forms, orbiters in the nuclear star cluster (NSC) will span a wide range of inclinations $(i)$, eccentricities $(e)$, orientations (prograde: $i<90^{\circ}$, or retrograde: $i>90^{\circ}$), and semi-major axes$(a)$. Once the AGN disc forms, some fraction of orbits will be coincident with the disc, yielding an initial embedded population, which can experience gas torques and migrate within the disc. The embedded population of stellar origin black holes (sBH) can encounter each other and merge yielding gravitational waves (GW) detectable with Advanced LIGO, Advanced Virgo and KAGRA \citep[e.g.][]{McKernan12,McKernan14,Bellovary16,Bartos17a,Stone17,Leigh18, McKernan18,Yang19,Tagawa20, Grobner20,Ford22,Vajpeyi22,Samsing22}. Stars embedded in AGN discs can rapidly grow in mass without shortening their lifetimes, as long as they continue to accrete fresh gas \citep{Cantiello20,Dittmann21,Jermyn22}.

Inclined orbiters not coincident with the disc ($i>\frac{H}{\rm R}$, the disc aspect ratio) experience drag when plunging through the disc, resulting in a systematic reduction of $(a,e)$, followed by inclination damping $(\frac{{\rm d}i}{{\rm d}t}<0)$ \citep{Rauch95,Just12,Kennedy16,Panamarev18,MacLeod20,Fabj20}. Thus, initially inclined orbiters will grow the embedded population in AGN discs over time, yielding larger potential sources of sBH-sBH mergers detectable with LVK, or sBH-SMBH mergers detectable with LISA. 

From \cite{Fabj20} (hereafter Paper~I), $\sim 10\%$ of prograde inclined orbiters $(i<90^{\circ})$ are captured by a \citet{Sirko03} type disc (hereafter SG) where density $\rho > 10^{-11}~{\rm g~cm^{-3}}$ for a plausible range of AGN disc lifetimes $(0.1 - 100~{\rm Myr})$, assuming accretion is negligible. Conversely, few prograde orbiters are captured by a lower gas density disc \citep[e.g.][]{Thompson05} (hereafter TQM). Paper~I also showed that most prograde stars that are captured have small initial $a<10^{3-4} \rm ~R_{g}$ which then decreases by at most an order of magnitude during capture, where ${\rm R_{g}} = G M_{\rm SMBH} ~c^{-2}$, the SMBH gravitational radius. Stellar-origin black holes (sBH) starting from a wide range of initial $a \leq 10^{6} \rm ~R_{g}$, end up at small disc radii $(a \la 10^{2} \rm ~R_{g})$ during capture. 

Paper I considered prograde orbits $(i<90^{\circ})$ only. However, we expect around half of all NSC orbits will have retrograde orientations $(i>90^{\circ})$ with respect to the gas disc, and most of these will not be coincident with the disc. So here we model the forces on inclined retrograde orbiters to determine the evolution of their orbital parameters over the lifetime of the disc, for a variety of orbiters and disc models. This paper is structured as follows. In \S\ref{sec:Methods} we summarize the methods and models from Paper~I. We describe our results in section \S\ref{sec:Results} and explore the wider impacts, notably on expected extreme mass ratio inspiral (EMRI) parameters and rates in section \S\ref{sec:Discussion}. Finally, in section \S\ref{sec:Conclusions} we summarize our conclusions.

\section{Models and Methods}\label{sec:Methods}
In this section we define drag torques at work for disc-crossing stellar and remnant orbiters (\S\ref{sec:Methods:Forces}), discuss the accretion disc models used in this study (\S\ref{sec:Methods:Models}), and outline our assumptions concerning nuclear star cluster populations (\S\ref{sec:Methods:NSC}). In \S\ref{sec:Methods:Capturetime} we define analytic approximations and find the results of numerical integration for capture-time, the time taken for a disc to capture an inclined orbiter.

\subsection{Forces on disc-crossing orbiters}\label{sec:Methods:Forces}
\begin{figure}\includegraphics[width=\linewidth]{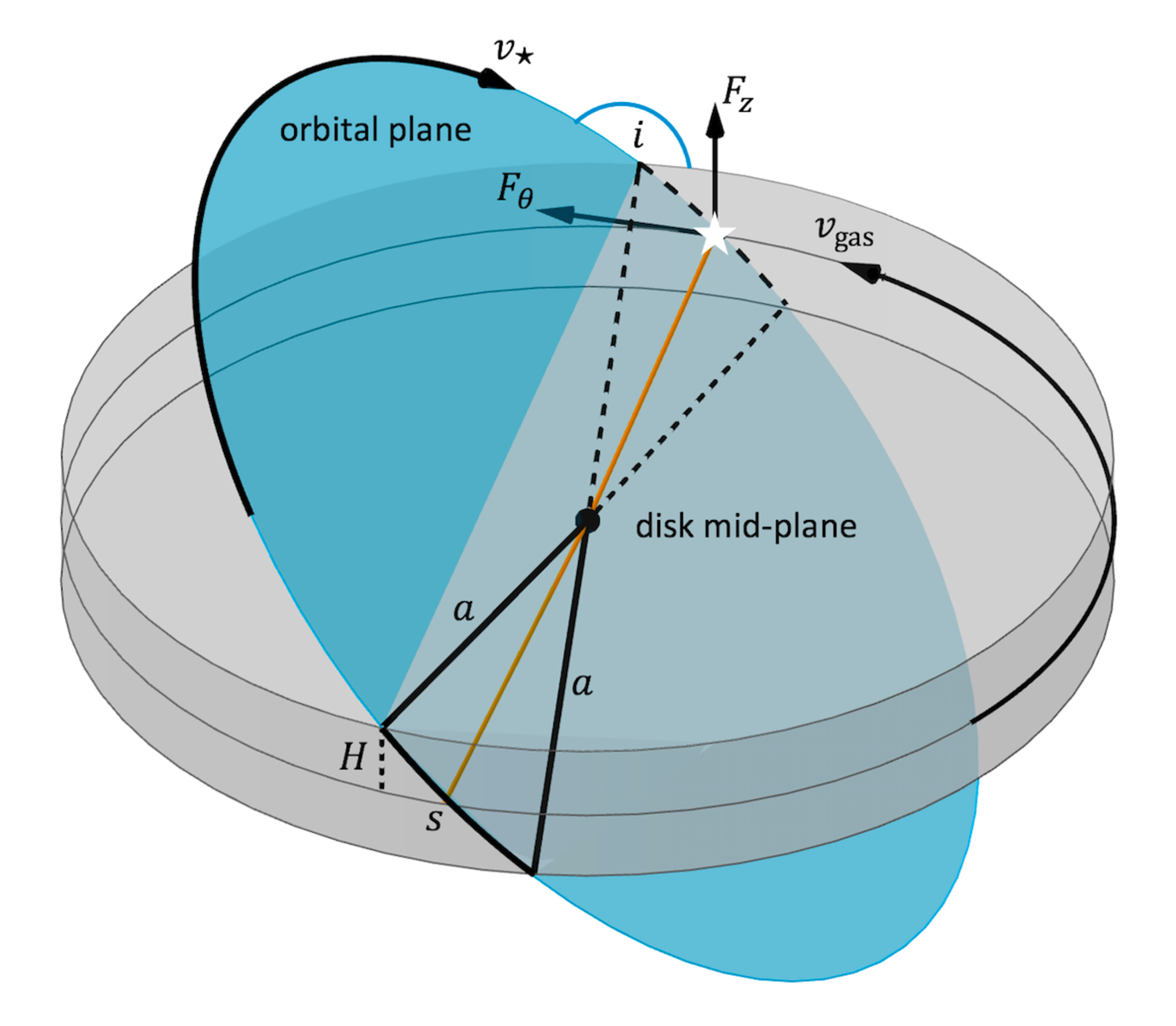}\caption{\label{fig:disccrossingorbiter} Orbital plane (blue) of a retrograde orbiter intersecting with the schematic disc of an AGN; the orange line represents the intersection of the disc mid-plane with the orbital plane. Semi-major axis $a$ is the distance from the central SMBH, scale height $H$ is half the total disc thickness, arc length $s$ represents the fraction of the orbit path crossing through the disc, $v_{\star}$ and $v_{\rm gas}$ are the Keplerian velocities of the orbiter and gaseous disc, and $F_{\rm z}$ and $F_{\rm \theta}$ are the vertical and angular components of the appropriate drag force, respectively.}\end{figure}

Fig.~\ref{fig:disccrossingorbiter} shows the intersection of a retrograde orbit plane intersecting with a prograde gas disc. Orbital inclination $i>90^{\circ}$ is retrograde and will yield a high relative velocity between the gas and the orbiter (the source of drag). Each orbital period consists of two distinct orbiter-disc interactions, where arc length $(s=2a~\arcsin{\frac{H}{a \sin{i}}})$\footnote{Note that we adjust the definition of scale height $H$ by a factor of two.} denotes the path through the disc for each interaction. $F_{\rm z}$ and $F_{\rm \theta}$ represent the vertical and angular components of the appropriate drag torque, and it is along path $s$ where these drag torques do work on the orbiter, yielding $\frac{{\rm d}i}{{\rm d}t}, \frac{{\rm d}a}{{\rm d}t}<0$ through repeated disc passages. Though orbital decay actively occurs during disc passage, we enforce circularity for each half orbit such that the net change in inclination $(\frac{{\rm d}i}{{\rm d}t})$ and semi-major axis $(\frac{{\rm d}a}{{\rm d}t})$ per passage takes effect after completion of each passage through the disc. Enforcing circularity means we also neglect the radial component of the drag torque, $F_{\rm r}$. Note also, following Paper~I, we assume negligible eccentricity $(e \sim 0)$ for inclined orbits, where the apocenter is outside of the disc, due to the eccentricity dampening and circularizing of elliptical orbits prior to the onset of inclination decay \citep{MacLeod20}. Note that it is possible that the eccentricity of retrograde orbiters is pumped as inclination is driven towards prograde. We ignore this effect for now, given that circularization happens for prograde inclinations. Nevertheless we should bear in mind that eccentricity pumping of retrograde orbiters that are ‘flipping’ orientation may drive them to smaller disc radii and therefore drive faster disc capture.

Following Paper~I, we apply geometric drag $({F}_{\rm GEO})$ to disc-crossing stars as \begin{equation}\label{eqn:Fgeo} {{F}_{\rm GEO} = \frac{1}{2} C_{\rm d} (4\pi r_{\ast}^{2}) \rho_{\rm disc} ~v^{2}_{\rm rel}} \end{equation} where drag coefficient $C_{\rm d} \approx 1$, $r_{\ast}$ is the stellar radius, $\rho_{\rm disc}$ is the local disc density, and $v_{\rm rel}$ is the relative velocity of the orbiter with respect to the disc. We apply Bondi Hoyle Lyttleton (BHL) drag $({F}_{\rm BHL})$ to disc-crossing sBH as \begin{equation}\label{eqn:Fbhl} {{F}_{\rm BHL} = 4 \pi ~G^{2} M_{\rm BH}^{2} ~\rho_{\rm disc} ~v^{-2}_{\rm rel}} \end{equation} where $M_{\rm BH}$ is the mass of the sBH. By taking into account the nature of the Mach cone trailing the orbiter we can estimate the dynamical drag as \citep{Ostriker99} \begin{equation}\label{eqn:Fdyn} {{F}_{\rm DYN} = 4 \pi ~G^{2} M_{\rm BH}^{2} ~\rho_{\rm disc} ~v^{-2}_{\rm rel} ~\ln\left( \frac{s}{r_{\ast}} \right).} \end{equation}

\begin{figure}\includegraphics[width=\linewidth]{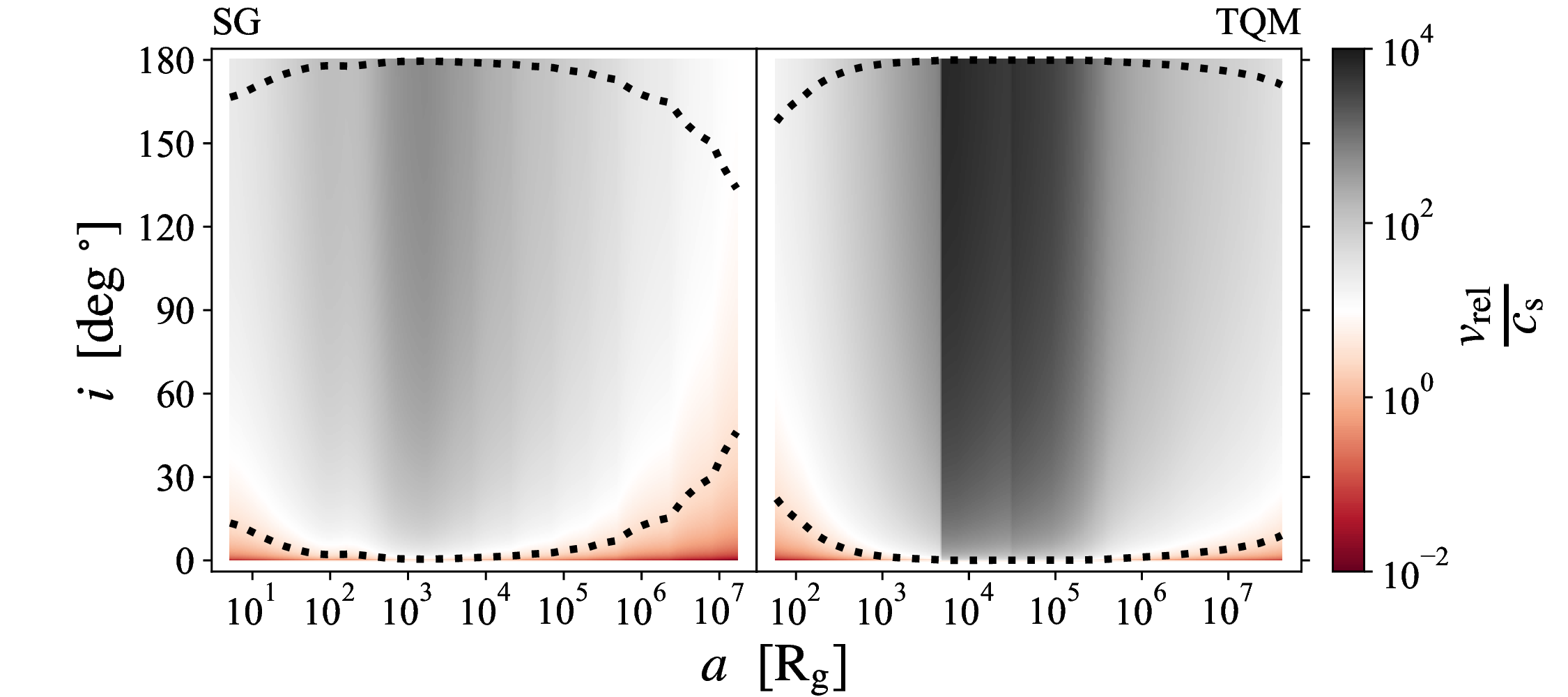}\caption{\label{fig:mach} Relative velocity to sound speed ratio $(\mathscr{M}=\frac{v_{\rm rel}}{c_{\rm s}})$ for a range of inclined orbit conditions with respect to SG (left) and TQM (right) accretion disc models. Dotted lines visualize the threshold, $i_{\rm crit}$ where orbiters $0^{\circ}<i<i_{\rm crit}<90^{\circ}$ are embedded on prograde orbits and orbiters $90^{\circ}<i_{\rm crit}<i<180^{\circ}$ are embedded on retrograde orbits. Red regions $(\mathscr{M}<4)$ correspond to orbit conditions where ${F}_{\rm DYN}>{F}_{\rm BHL}$. Grey regions $(\mathscr{M}>4)$ correspond to orbit conditions where ${F}_{\rm BHL}>{F}_{\rm DYN}$. Conditions favoring dynamical friction mainly coincide with embedded prograde orbits.}\end{figure}

In Fig.~\ref{fig:mach} we show $\mathscr{M}=\frac{v_{\rm rel}}{c_{\rm s}}$, the relative velocity to sound speed ratio, as a function of $(a,i)$ for the SG and TQM disc models considered here. For most regions of the disc $(a)$ and most inclinations $(i)$, the inclined orbiters are highly supersonic $(\mathscr{M}\gg 1)$, in grey. The only exceptions (in red) are at low prograde inclinations at relatively small $a$, where capture-time is already expected to be quite rapid, or relatively large $a$ for SG type discs. As a result, we neglect the effects of dynamical drag in our analysis, particularly for retrograde orbiters. We also neglect any gravitational torques on the orbits, as the disc mass enclosed is small compared to the SMBH mass, and any precessional or dynamical effects due to the gravitation of the disc will be small compared to the drag forces we consider.

\subsection{AGN disc models}\label{sec:Methods:Models}
\begin{figure}\includegraphics[width=\linewidth]{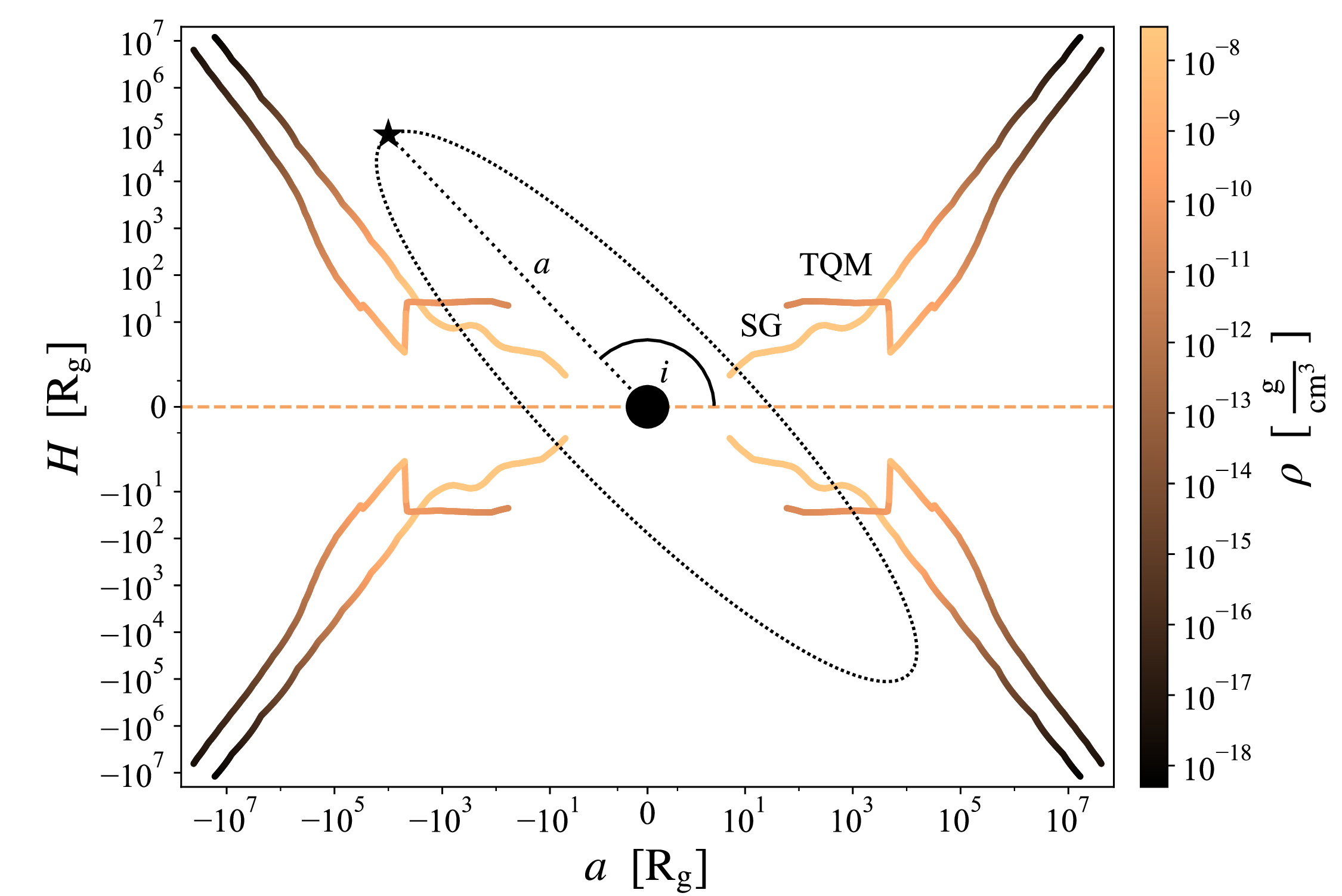}\caption{\label{fig:discprofiles} 1-D scale height $(H)$ and density $(\rho)$ profiles for SG and TQM accretion disc models, scaled, reflected and centered around a $10^{8}$~\msun~SMBH. The schematic inclined orbit is defined by its inclination $(i)$ and semi-major axis $(a)$. We assume negligible eccentricity $(e \sim 0)$ for inclined orbits.}\end{figure}

Following Paper~I, we use \citet{Sirko03} and \citet{Thompson05} as our fiducial AGN disc models. SG was designed to match the spectral energy distribution of AGN in the optical/UV (dominated by the inner disc) but assumes that the outer disc is heated in an unspecified manner to prevent collapse. So SG is a reasonable approximation for dense inner discs, but is not a reliable model of outer discs. Conversely, TQM was designed to match large-scale mass accretion rates in galactic nuclei, and represents a more plausible outer disc model but is probably less reliable in the inner disc. Fig.~\ref{fig:discprofiles} shows the 1-d radial profiles of both models\footnote{Note that TQM exhibits sharp changes as $a \to 5 \times 10^{3} \rm ~R_{g}$ due to their choice of opacity model; however, this discontinuity may be eliminated by a different choice of opacities (e.g. \citealt{Dittmann20}).}, where the colour bar illustrates the higher density inner disc (orange) and lower density outer disc (black). Paper~I showed that disc gas density $\rho \geq 10^{-11} {\rm~g~cm^{-3}}$ (i.e the inner regions of both disc models) was most effective at accelerating disc capture.

\subsection{Nuclear Star Clusters}\label{sec:Methods:NSC}
\begin{deluxetable}{lcc}\tablecolumns{3}\tablewidth{170pt}\tablecaption{Assumed Properties of NSC Orbiters}\tablehead{ \colhead{Object} & \colhead{Mass~(\msun)} & \colhead{Radius~(\rsun)} } \startdata O~Star & 50 & 15\\ G~Star & 1 & 1\\ M~Dwarf & 0.5 & 0.4\\ sBH & 10 & --\\ sBH & 50 & --\\ \enddata \label{table:properties}\end{deluxetable}

Nuclear Star Clusters (NSC) provide the initial population of stars and stellar remnants that may be captured by AGN discs. NSC are the densest star clusters in the local Universe \citep{Boker02,Boker04,Leigh12,scott13,Antonini13,Georgiev14,Antonini15,Georgiev16,Neumayer20}. NSC are similar in size to globular clusters (few pc across) but are typically more massive \citep[see e.g.][for more details]{Leigh15,Antonini15}. NSC are preferentially detected around lower mass SMBH, and observed stellar light distributions indicate a decreasing \textit{fractional} contribution from stars to the total nuclear luminosity with increasing SMBH mass \citep[see e.g.][and references therein]{Neumayer20}. Nevertheless, stars and their remnants are still expected in galactic nuclei at all SMBH masses; the point is simply that the same mass of stars in a NSC around a $10^{6}$~\msun~SMBH is far more difficult to detect around a $10^{8}$~\msun~SMBH. Here we assume that the processes bringing stars and their remnants into galactic nuclei (e.g. major and minor mergers, dynamical friction acting on clusters, star formation) operate around all mass SMBH and can produce similar quantities of stars.

As in Paper~I, we assume a $10^{8}$~\msun~ central SMBH in the center of our AGN. We follow the orbital evolution of three star-types (O~Stars, G~Stars, and M~Dwarfs) and sBH of mass 10~\msun~ and 50~\msun~ (listed in Table~\ref{table:properties}). We ignore changes in stellar evolution due to the interaction with (and accretion from) the disc \citep[see e.g.][for modest, and AGN-like gas density accretion, respectively]{Leigh16,Cantiello20}. However, further theoretical work to constrain these effects is highly desirable.

\subsection{Constraining capture-time}\label{sec:Methods:Capturetime}
Through repeated passage, orbit parameters $(a,i)$ evolve over capture-time $(T_{\rm cap})$ until the orbiter either ends up at the inner edge of the disc model $(a=a_{\rm min})$, or $i \approx i_{\rm crit}$ such that the entire path of its orbit is embedded within the disc. Dotted lines in Fig.~\ref{fig:mach} show $i_{\rm crit}~(a)$ for SG and TQM models respectively.

We use two different methods for estimating $T_{\rm cap}$. First, we assume that the change in semi-major axis $a$ is negligible $(\frac{{\rm d}a}{{\rm d}t} \approx 0)$ and calculate $T_{\rm cap}$ only at $a_{\rm o}$; an easily calculable upper limit to $T_{\rm cap}$. Second, we calculate $T_{\rm cap}$ by numerically integrating over many orbits, iterating $\frac{{\rm d}a}{{\rm d}t}$ per disc interaction due to drag torques. Paper~I shows that the analytic approximations given by \begin{equation}\label{eqn:Tgeo} {T_{\rm GEO} ~\la ~\frac{8 ~r_{\ast} ~\rho_{\ast}}{3\pi s ~\rho_{\rm disc} ~\sin i} ~T_{\rm orb}} \end{equation} \begin{equation}\label{eqn:Tbhl} {T_{\rm BHL} ~\la ~\frac{v_{\rm rel}^{4} ~\sin^{3}i}{4\pi^{2} s ~\rho_{\rm disc} ~G^{2} M_{\rm BH}} ~T_{\rm orb}} \end{equation} are useful upper limits to $T_{\rm cap}$ for geometric drag and BHL drag respectively. We take the same approach to numerical integration as in Paper~I, accounting for the work done twice per orbit (i.e at each disc passage): \begin{equation}\label{eqn:Tcap} {T_{\rm cap} ~\approx ~\sum_{\rm j=0}^{2N_{\rm cap}} ~\frac{\pi ~a_{\rm j+1}^{3/2}}{\sqrt{G M_{\rm SMBH}}} ~= ~\sum_{\rm j=0}^{2N_{\rm cap}} ~\frac{T_{\rm orb}~(a_{\rm j})}{2}.} \end{equation} Computational efficiency for a given numerical simulation therefore favors low $N_{\rm cap}~(i_{\rm o},a_{\rm o})$, the number of orbits over which $T_{\rm cap}~(i_{\rm o},a_{\rm o})$ is integrated.

\section{Results}\label{sec:Results}
Here we discuss the evolution (\S\ref{sec:Results:Evolution}) and capture-time (\S\ref{sec:Results:Capturetime}) of inclined orbits caused by drag torques on stars and sBH as they interact with SG and TQM accretion disc models. We present numerical integration results for a full range of initial inclination angles $(i_{\rm o})$, including prograde $(0^{\circ}<i_{\rm o}<90^{\circ})$ and retrograde $(90^{\circ}<i_{\rm o}<180^{\circ})$. 

\subsection{Evolution of inclined orbits}\label{sec:Results:Evolution}
\begin{figure}\includegraphics[width=\linewidth]{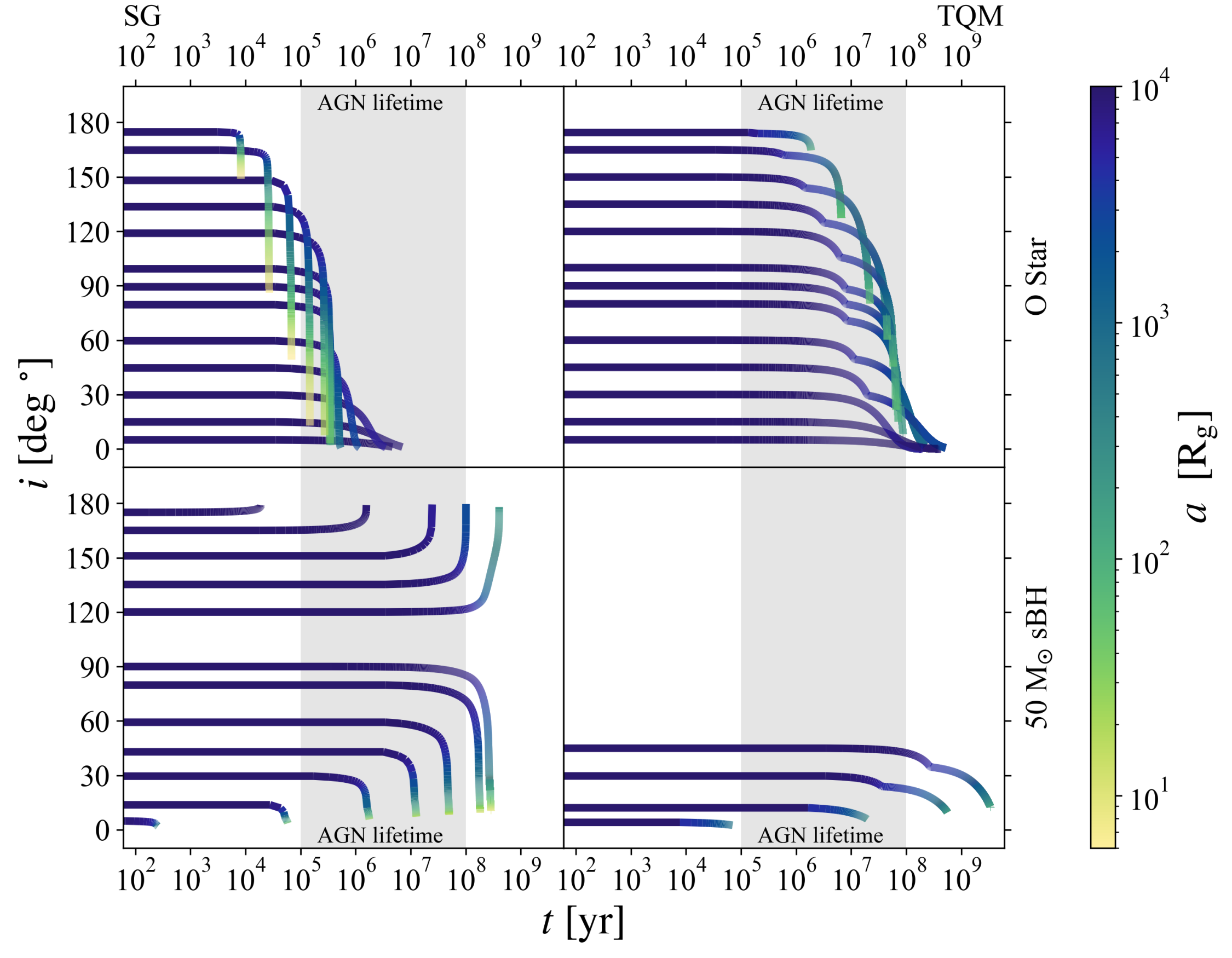}\caption{\label{fig:evolution} Evolution of $(i,a)$ during $T_{\rm cap}$ for O~Stars (top) and 50~\msun~sBH (bottom) interacting with SG (left) and TQM (right) discs. Orbit evolution for other star-types and sBH masses differ only in $T_{\rm cap}$. Grey bands represent a plausible range of AGN lifetime $(\tau_{\rm AGN} \sim 0.1 - 100~{\rm Myr})$. Evolution of curves progress from left to right. Changes in colour along a curve indicates change in $a$ during capture from $a_{\rm o}=10^{4} \rm ~R_{g}$ to $\sim 10^{2} \rm ~R_{g}$ (green) and $\sim 10 \rm ~R_{g}$ (yellow). Vertical changes in curves indicate a change in $i$ from $i_{\rm o}$. Strong vertical changes indicate a runaway capture in a short time. The bump in TQM curves are a relic of the unphysical jump in the disc height and density profiles near $5 \times 10^{3} \rm ~R_{g}$. Note that disc capture for many initially retrograde stars is terminated prematurely $(i \gg 0^{\circ})$, because $\Delta a$ is so large that the orbiter essentially decays directly onto the SMBH (yellow colour on curves). During their orbit evolution, stars in retrograde orbits are always flipped towards prograde orientation, while sBH at $i_{\rm o}> 113^{\circ}$ are driven towards fully retrograde capture.}\end{figure}

Fig.~\ref{fig:evolution} shows orbital $(i,a)$ evolution over time for O~Stars (top) and 50~\msun~sBH (bottom), crossing SG (left) and TQM discs (right) given a range of initial inclinations, $i_{\rm o}$ but $a_{\rm o} = 10^{4} \rm ~R_{g}$. Vertical changes in a curve indicates a change in $i$ and the colour change on a given curve shows the change in $a$. The grey shaded region corresponds to a plausible range of AGN disc lifetimes $(\tau_{\rm AGN} \sim 0.1 - 100~{\rm Myr})$. Orbital evolution for other star-types and sBH masses have the same form but different $T_{\rm cap}$.

For stars (top), since ${F}_{\rm GEO} \propto v_{\rm rel}^{2}$, retrograde stars experience a stronger headwind and therefore shorter $T_{\rm cap}$ and larger $\Delta a$ than prograde stars. For SG discs (top left) as $i_{\rm o}$ increases from $i \sim 0^{\circ}$, $T_{\rm cap}$ decreases rapidly from $\sim 10~$Myr, stabilizes around $\sim 1~$Myr for polar $i_{\rm o} \sim 90^{\circ}$ and then decreases rapidly for increasingly retrograde $i_{\rm o}$. ${F}_{\rm GEO}$ also acts to push retrograde orbits towards prograde inclinations on the way to disc capture, although some initially retrograde orbiters $(i \sim 180^{\circ})$ are still captured in retrograde orientation. In some initially retrograde cases, $a$ decays to the innermost limits of the accretion disc model before reaching critical inclination, presumably resulting in accretion onto the central SMBH rather than being captured by the accretion disc itself. We can identify these cases in Fig.~\ref{fig:evolution} as those whose final inclinations at capture are greater than the $i_{\rm crit}$ corresponding to their semi-major axis at capture (see dotted lines in Fig.~\ref{fig:mach}). 

In the case of sBH, the main result is that for $i_{\rm o} >113^{\circ}$, sBH are driven towards retrograde capture. Disc capture-time $(T_{\rm cap})$ is longest for polar or near polar initial orbits $(i_{\rm o} \sim 90^{\circ})$ and shortest for orbits closest to embedded prograde $(i_{\rm o} \sim 0^{\circ})$ or embedded retrograde $(i_{\rm o} \sim 180^{\circ})$. There is less change in semi-major axis $(\Delta a)$ during capture for sBH compared to stars and $\Delta a$ is smaller for retrograde sBH than prograde sBH.

Furthermore, unlike the case for stars, we find that inclined sBH orbits always decay to be fully embedded in the disc before the sBH reach the innermost limits of the accretion disc. As a result sBH always experience disc capture before SMBH capture. Hence, all EMRI generated by interactions with accretion discs should have $i=0^{\circ}$ or $180^{\circ}$.

\subsection{Capture-time for inclined orbits}\label{sec:Results:Capturetime}
\begin{figure}\includegraphics[width=\linewidth]{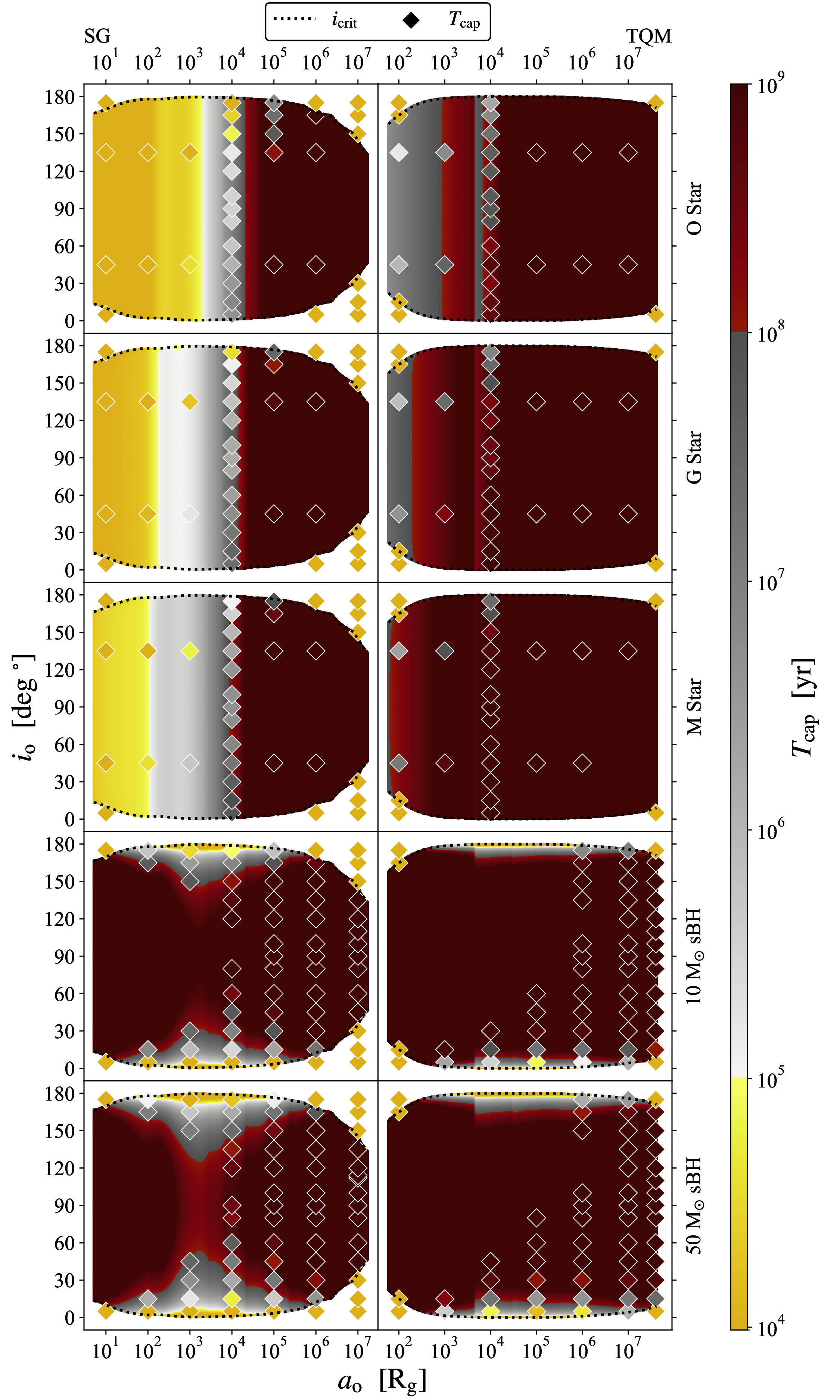}\caption{\label{fig:capturetime} Capture-time, $T_{\rm cap}$~[Myr], for a range of initial conditions $(a_{\rm o},i_{\rm o})$ and object types (O,~G,~M~stars and 10~\msun,~50~\msun~sBH). Colours indicate $T_{\rm cap}$ relative to AGN lifetime $(\tau_{\rm AGN})$: yellow $(T_{\rm cap}<\tau_{\rm AGN})$, grey $(T_{\rm cap} \sim \tau_{\rm AGN})$, red $(T_{\rm cap} > \tau_{\rm AGN})$. Background colour indicates $T_{\rm cap}=T_{\rm GEO}~(T_{\rm BHL})$ for stars~(sBH) respectively and shows the analytic \textit{upper limit} for $T_{\rm cap}$. Diamond symbols indicate the result of direct numerical integration for that $(a_{\rm o},i_{\rm o})$. Where a diamond symbol contains a different colour from background, numerical integration is revealing $T_{\rm cap} \ll T_{\rm GEO,BHL}$ upper limits. Dotted curves show the threshold, $i=i_{\rm crit}$, at which point the retrograde (top dotted curve) or prograde (bottom dotted curve) orbiter is captured by the disc.}\end{figure}

Fig.~\ref{fig:capturetime} shows how long it takes for a disc to capture an inclined orbit $(T_{\rm cap})$~[Myr] for a range of initial inclination angles $(i_{\rm o})$ and initial semi-major axes $(a_{\rm o})$ for O,~G,~M~stars and 10~\msun,~50~\msun~sBH. Dotted curves indicate prograde~(bottom) and retrograde~(top) $i_{\rm crit}$ thresholds, representing critical inclinations for disc capture. For plausible AGN lifetimes $(\tau_{\rm AGN} \sim 0.1 - 100~{\rm Myr})$, colours indicate $T_{\rm cap}<\tau_{\rm AGN}$ (yellow), $T_{\rm cap} \sim \tau_{\rm AGN}$ (grey) and $T_{\rm cap}>\tau_{\rm AGN}$ (red). Diamond symbols correspond to specific numerical integration results for particular choices of $(a_{\rm o},i_{\rm o})$ and background colours correspond to upper limit to $T_{\rm cap}$ given by $T_{\rm GEO,(BHL)}$ for stars(sBH) respectively. Where diamond symbols are coloured differently from the background upper limit, numerical integration is showing that $T_{\rm cap} \ll T_{\rm GEO,BHL}$. 

In general, Fig.~\ref{fig:capturetime} shows that all stars with $a_{\rm o} \leq 10^{4} \rm ~R_{g}$ (whether retrograde or prograde) are rapidly captured within $\tau_{\rm AGN}$ by SG-like AGN discs. Stars beyond this orbit are not captured. For TQM-like discs, only a small fraction of stellar orbits within $a_{\rm o} \leq 10^{4} \rm ~R_{g}$ are captured by the disc, and preferentially the more massive stars. A significant fraction of sBH orbits within $a_{\rm o} \sim 10^{5} \rm ~R_{g}$ and $i>150^{\circ}; ~i<30^{\circ}$ are captured by SG-like discs. For TQM-like discs, the sBH within $i>165^{\circ}; ~i<15^{\circ}$ are plausibly captured across a wide range of disc radii.

\subsubsection{Stellar capture-times}\label{sec:Results:CapturetimeST}
\begin{figure*}\includegraphics[width=\linewidth]{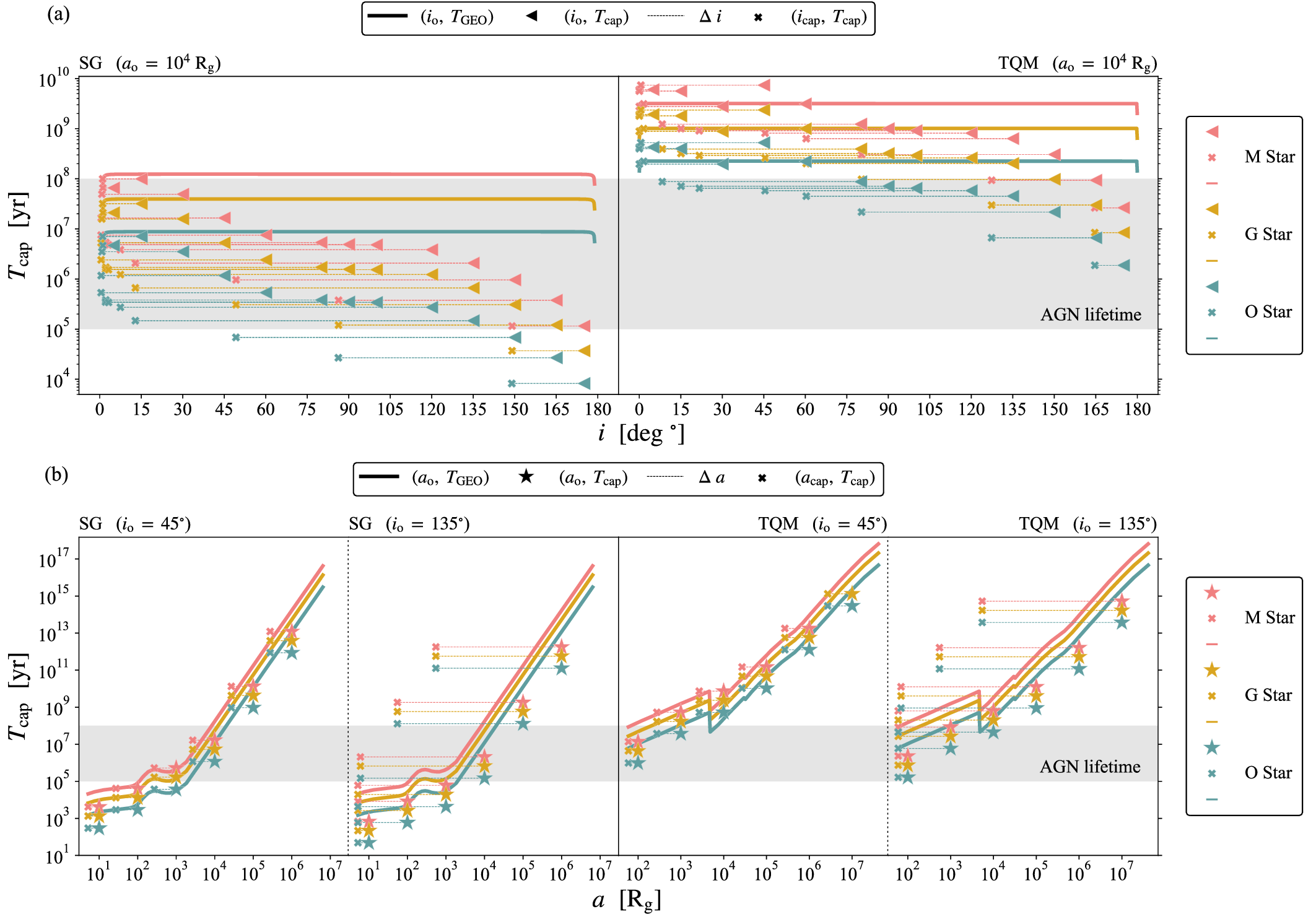}\caption{\label{fig:capturetimeST} {\bf (a)} $T_{\rm cap}~(i_{\rm o})$, for O~(teal), G~(gold), M stars~(red) interacting with SG~(left) and TQM~(right) disc models assuming $a_{\rm o}=10^{4} \rm ~R_{g}$ in all cases. Grey~bands indicate a plausible range of AGN lifetimes $(\tau_{\rm AGN} \sim 0.1 - 100~{\rm Myr})$. Solid lines correspond to the upper limit $T_{\rm cap}=T_{\rm GEO}~(a_{\rm o},i_{\rm o})$. Symbols (triangles and xs) correspond to initial and final inclination for a given $T_{\rm cap}$ numerical integration (indicated by dashed line). {\bf (b)} $T_{\rm cap}~(a_{\rm o})$ for the same stars in (a) but assuming $i_{\rm o}=45^\circ~\&~135^\circ$. Solid lines are as in (a). Symbols (stars and xs) correspond to initial and final semi-major axis for a given $T_{\rm cap}$ numerical integration (indicated by dashed line). At $10^{4} \rm ~R_{g}$, for the SG model all three stellar types are captured within $\tau_{\rm AGN}$ regardless of their initial orientation, while for TQM only most massive stars at a retrograde orientation can be captured within $\tau_{\rm AGN}$.}\end{figure*}

Fig.~\ref{fig:capturetimeST}~(a), shows $T_{\rm cap}$ for a range of $i_{\rm o}$ (given $a_{\rm o}=10^{4} \rm ~R_{g})$ for three representative star-types: O~Stars (teal), G~Stars (gold), and M~Dwarfs (orange) interacting with SG (left) and TQM (right) discs. Fig.~\ref{fig:capturetimeST}~(b), shows $T_{\rm cap}$ for a range of $a_{\rm o}$ (given $i_{\rm o}=45^{\circ},135^{\circ})$ for the same stars and discs. Grey shaded areas represent a plausible range of AGN lifetimes $(\tau_{\rm AGN})$ as above. Solid lines represent the upper limit to $T_{\rm cap}$ for each stellar type given by $T_{\rm GEO}$. In Fig.~\ref{fig:capturetimeST}~(a)~\&~(b) triangle and star symbols respectively indicate starting values of $i_{\rm o}$ $(a_{\rm o})$ and x symbols indicate inclination (semi-major axis) at capture after numerical integration, with dashed lines depicting $\Delta i$ $(\Delta a)$ for that integration. Where a given integration in Fig.~\ref{fig:capturetimeST}~(a) does not end around $i\sim 0^{\circ}$, this indicates that the final semi-major axis is very close to the SMBH.

From Fig.~\ref{fig:capturetimeST}~(a), most stars within $a_{\rm o} \leq 10^{4} \rm ~R_{g}$ can be captured by an SG-like disc. Retrograde orbiters are captured fastest as are more massive stars. Inclined orbits are captured significantly faster than the upper limit $(T_{\rm cap} \sim T_{\rm GEO})$ because $\frac{{\rm d}a}{{\rm d}t}<0$ and a runaway drag effect due to increased number of passes through an increasingly dense disc. For TQM-like discs, the same conclusions hold, but the timescales are longer due to lower disc density. Thus, only retrograde massive stars are likely captured by a TQM-like disc.

From Fig.~\ref{fig:capturetimeST} (b), prograde stars decrease $a$ by at most an order of magnitude during capture but retrograde stars experience much greater changes in semi-major axis during capture due to the much larger headwind. Retrograde stars with $i_{\rm o} \geq 135^{\circ}$ at $a_{\rm o} \leq 10^{4~(3)} \rm ~R_{g}$ end up on the SMBH within $\tau_{\rm AGN}$ for SG~(TQM)~--like discs. 

\subsubsection{sBH capture-times}\label{sec:Results:CapturetimeBH}
\begin{figure*}\includegraphics[width=\linewidth]{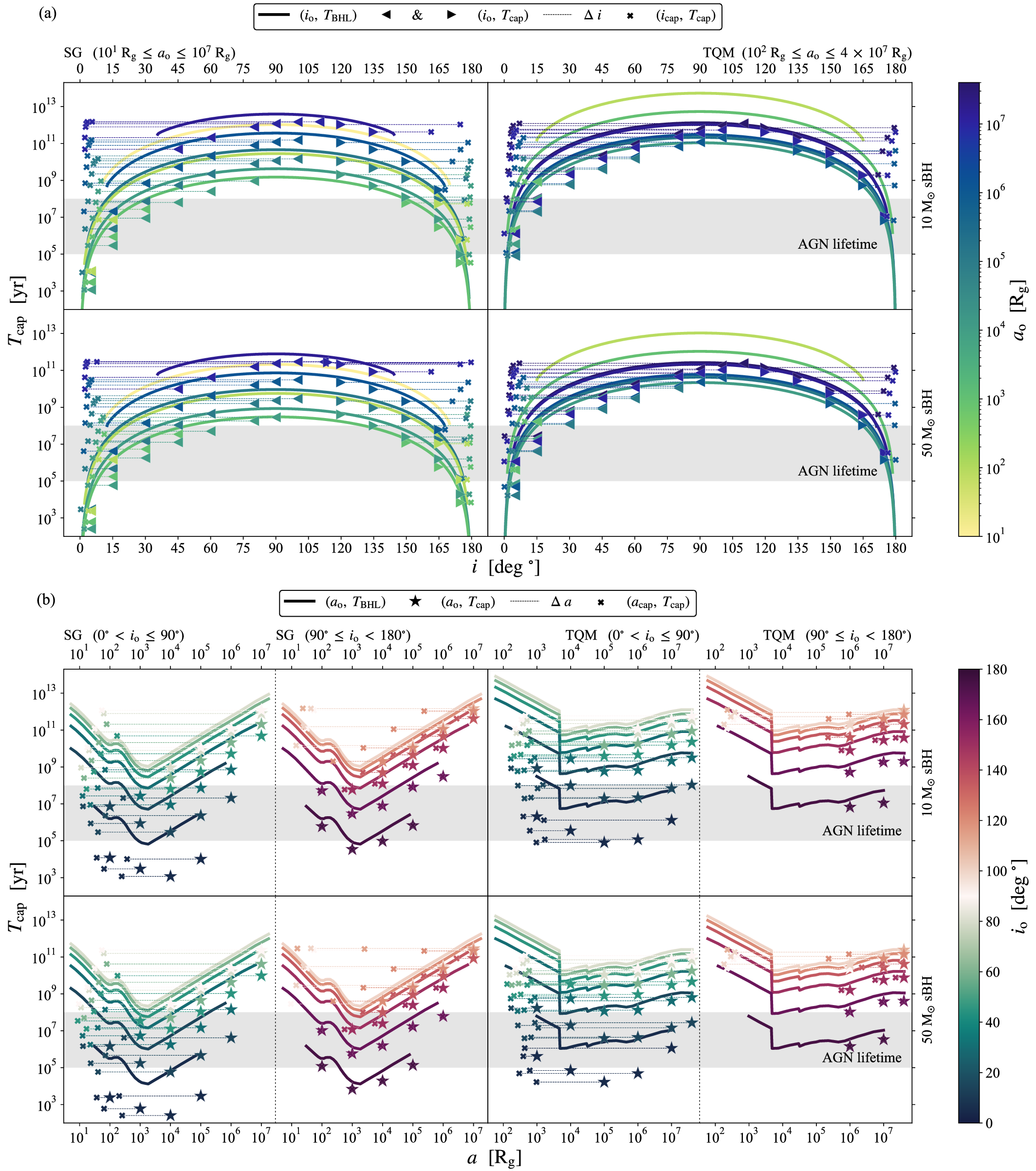}\caption{\label{fig:capturetimeBH} {\bf (a)} $T_{\rm cap}~(i_{\rm o})$, for 10~\msun~sBH~(top) and 50~\msun~sBH~(bottom) interacting with SG discs~(left) and TQM discs~(right). Grey~bands as in Fig.~\ref{fig:capturetimeST}. Colours indicate $a_{\rm o}~({\rm R_{g}})$. Solid lines correspond to the upper limit $T_{\rm cap}=T_{\rm BHL}~(a_{\rm o},i_{\rm o})$. Triangle-symbols, x-symbols, and dashed lines are as in Fig.~\ref{fig:capturetimeST}~(a). Left-pointing triangle symbols indicate $\frac{{\rm d}i}{{\rm d}t}<0$ towards $i \sim 0^{\circ}$. Right-pointing triangle symbols indicate $\frac{{\rm d}i}{{\rm d}t}>0$ towards $i \sim 180^{\circ}$. {\bf (b)} $T_{\rm cap}~(a_{\rm o})$ for the same sBH as in (a). Colours now indicate $i_{\rm o}~( ^{\circ})$. Solid lines are as in (a). Star-symbols, x-symbols, and dashed lines are as defined in Fig.~\ref{fig:capturetimeST}~(b). For both disc models, orbiters at nearly fully prograde $(i \leq 15^{\circ})$ and retrograde $(i \geq 165^{\circ})$ initial inclinations experience the fastest capture-times. Under similar initial conditions, on average higher mass sBH experience a shorter $T_{\rm cap}$.}\end{figure*}

Fig.~\ref{fig:capturetimeBH} depicts (a) $T_{\rm cap}~(i_{\rm o})$ and (b) $T_{\rm cap}~(a_{\rm o})$ for sBH of 10~\msun~(top) and 50~\msun~(bottom) interacting with SG~(left) and TQM~(right) discs. As in Fig.~\ref{fig:capturetimeST}, solid lines indicate upper limits, this time $T_{\rm cap}=T_{\rm BHL}$, coloured according to: (a) $a_{\rm o}$ or (b) $i_{\rm o}$. Triangle and Star Symbols indicate the initial values and x symbols indicate the end values of each numerical integration (shown by dashed lines). Triangle symbols pointing leftward~(rightward) in Fig.~\ref{fig:capturetimeBH}~(a) correspond to $\frac{{\rm d}i}{{\rm d}t}<0~(>0)$ as $i \to 0^{\circ}~(180^{\circ})$. Grey shaded regions are as in Fig.~\ref{fig:capturetimeST}. 

From Fig.~\ref{fig:capturetimeBH}~(a) we can see that nearly fully prograde $(i \leq 15^{\circ})$ or nearly fully retrograde $(i \geq 165^{\circ})$ sBH are generally captured fastest by both SG- and TQM-like discs. Near-polar orbits are captured last (if at all). For SG-like discs, sBH from much of the spherical component $(i_{\rm o} \geq 140^{\circ}, ~i_{\rm o} \leq 60^{\circ})$ can actually be captured around the densest part of the disc $(a_{\rm o} \sim 10^{3} \rm ~R_{g})$ within a fiducial AGN lifetime. TQM-like discs are lower density than SG-like discs in general with density maxima at $a_{\rm o} \sim 10^{4} \rm ~R_{g}$ so disc capture takes longer and fewer inclined orbiters are captured overall. From Fig.~\ref{fig:capturetimeBH}~(a) there is a critical $i_{\rm ret} \sim 113^{\circ}$ below~(above) which orbiters are driven to $i \to 0^{\circ}~(180^{\circ})$ (in contrast to \citealt{Generozov23}). It is unclear whether $i_{\rm ret}$ is a universal critical angle, since some physics may be missing in our treatment (e.g. we consider exclusively circular orbits and we neglect the disc back-reaction). We shall investigate this effect in future work. Fig.~\ref{fig:capturetimeBH}~(b) confirms the general trends of Fig.~\ref{fig:capturetimeBH}~(a), but reveals two interesting properties: First, that $\Delta a$ is far larger for prograde capture than retrograde capture for both disc types. Second, that the upper limits to $T_{\rm cap}$ are a reasonable approximation for retrograde sBH capture, but a significant over-estimate for low inclination prograde sBH. Fig.~\ref{fig:capturetimeBH}~(b) also shows that more massive sBH are typically captured faster, as we should expect.

As orbiters decrease $a$ during the process of capture, both stars and sBH face more intense drag torques due to (i) typically denser inner disc and (ii) higher orbital frequencies. Technically, even though $T_{\rm cap}$ decreases, the larger number of disc passages results in longer processing times for numerical integration, as each disc passage requires processor cycles (and cannot be readily parallelized). This effect can be seen in the absence of diamond points (numerical integrations) at small values of $a_{\rm o}~({\rm R_{g}})$ in Fig~\ref{fig:capturetime}. 

\section{Discussion}\label{sec:Discussion}
Our results have important implications for expectations of LVK GW detections of binary black hole (BBH) mergers, especially at high mass (\S\ref{sec:Discussion:BBH}), LISA detections of EMRI (\S\ref{sec:Discussion:EMRI}), the presence of tidal disruption events (TDE) and other electromagnetic observables in AGN (\S\ref{sec:Discussion:TDE}), and the evolution of stars in NSC, especially for nuclei in a post-AGN state (\S\ref{sec:Discussion:RG}).

\subsection{Ground-based GW observables: BBH mergers}\label{sec:Discussion:BBH}
Our results show that denser SG-like discs are more efficient at sBH capture than TQM-like discs. More massive sBH are also generally captured more easily than less massive sBH. From Fig.~\ref{fig:capturetime}, for a spherical distribution of 10~(50)~\msun~sBH orbiters interacting with an SG-like disc, perhaps as much as $\frac{1}{6}~(\frac{1}{3})$ could be captured within $\tau_{\rm AGN} \sim 100~{\rm Myr}$. The corresponding fraction for TQM-like discs is $<5\%$ of sBH orbits captured.

If we assume there are $\mathcal{O} (10^{4})$~sBH in the inner cubic parsec of an NSC, uniformly distributed in $\log (a)$, and we ignore mass segregation, or an sBH preferential plane, then for an SMBH of $10^{8}$~\msun~ with a SG disc, we might expect 10~(30)\% of 50~\msun~sBH or 7~(25)\% of 10~\msun~sBH captured within $\tau_{\rm AGN} \sim ~1~(100)$~Myr disc lifetime. This would imply up to $\mathcal{O} (10^{3})$~sBH embedded in pc-scale AGN accretion discs for modest $\tau_{\rm AGN}$. However an SG-like disc around SgrA$\ast$ extending to $\sim 0.1$pc captures only $\mathcal{O} (10^{2})$~sBH. Quite simply, large-scale dense AGN discs (like SG) should contain a large population of sBH and small-scale AGN discs should contain a small embedded population. The embedded population of lower density discs (like TQM) will tend to be dominated by the fraction of the NSC that have orbits coincident with the disc and only a small admixture will arrive over time from disc capture.

The higher the disc density, the higher the migration rate of embedded objects. If dense AGN discs contain migration traps \citep{Bellovary16} then a pile-up of migrators can occur and complex dynamics (but also mergers) should result \citep{Secunda19,Secunda20,McKernan19a}, although mergers can also occur at a high rate in the bulk disc away from the trap \citep{Tagawa20,McKernan20}. In general therefore, we expect the BBH merger rate is highest, in denser, large-radius AGN discs of moderate lifetime. This simple picture is complicated by the fact that sBH on retrograde orbits can be efficiently captured by dense AGN discs. Retrograde sBH embedded in the disc may act as efficient ionizers of BBH during scattering encounters. However, retrograde sBH should rapidly experience eccentricity pumping and orbital decay to very small disc radii \citep{Secunda21}, effectively segregating this population close to the SMBH.

On average, for similar initial conditions, larger $M_{\rm BH}$ correspond to shorter $T_{\rm cap}$. We therefore expect higher mass sBH to spend more time in AGN discs than lower mass sBH. This implies that more massive sBH in AGN should be preferentially torqued towards alignment with the disc gas via accretion, whereas less massive sBH, which take longer to be captured by the disc, should have a more random distribution of spin orientations. Such a distribution of spin orientation with sBH mass may account for the intriguing anti-correlation observed by LVK in BBH mass ratio and effective spin \citep{Callister21,McKernan22,Wang21}.

\subsection{AGN EMRI}\label{sec:Discussion:EMRI}
An EMRI occurs if an sBH $(M_{\rm BH}<10^{-4}~M_{\rm SMBH})$ at small $a_{\rm o}$ gradually decays onto the SMBH, emitting gravitational waves. EMRI resulting from two-body scattering of sBH within the NSC into EMRI orbits are well studied \citep[e.g.][]{Sigurdsson97,Alexander05,Merritt06}. However if AGN discs can efficiently capture sBH at small disc radii, as outlined above, then there must be a \textit{new} potential source of EMRI detectable with LISA (AGN-EMRI; also pointed out in Paper~I).

The details of what happens to the population of sBH at small radii in AGN are complex and far beyond the scope of this paper. Nevertheless, in the first $\sim 0.1~$Myr of an SG-like AGN, we expect fully embedded retrograde sBH at radii $a\leq 10^{3} \rm ~R_{g}$ experience rapid eccentricity pumping and semi-major axis decay \citep{Secunda21}. Many of these retrograde sBH never form LISA-detectable EMRI as they merge eccentrically with the SMBH. However, $\geq~50$~\msun~ retrograde sBH can circularize at small radii \citep{Secunda21}, and this population of sBH should dominate the AGN-EMRI population early on $(\leq 0.1~{\rm Myr})$. Fig.~\ref{fig:capturetimeBH}~(b) shows a small admixture of disc-skimming prograde sBH also captured at small radii during this time but we expect large-angle scatterings \citep[e.g.][]{Wang21} remove prograde sBH either to large radii or onto the SMBH. Since retrograde sBH experience far less efficient gas drag than prograde sBH, we expect that signatures of gas drag will not be discernible among this initial population of AGN-EMRI.

After this first phase of the AGN, we expect the prograde population to grow at small radii via either disc capture or migration within the disc. If migration traps \citep{Bellovary16} are not common to AGN discs, then fully embedded prograde sBH should eventually migrate to small disc radii and remove retrograde sBH via large angle scatterings during dynamical encounters. Thus, the detection of the effects of gas drag among the population of LISA EMRI would imply both that AGN-EMRI can occur and that migration traps may not be common in AGN. Note that all the above applies to SG-like discs. If AGN discs are more like TQM, then the same processes apply, but the timescales involved increase considerably due to lower gas densities.

In general, LISA should observe AGN-EMRI which are coplanar with the AGN disc, with a constant inclination. An aligned retrograde but eccentric sBH would instead merge with the SMBH on a highly eccentric orbit (resulting in a different but unique gravitational wave signal \citep{Secunda21}). We predict that a subset of EMRI detectable with LISA (AGN-EMRI) will show circular, planar orbits, and that these orbital signatures are a result of AGN disc capture and migration. In fact, {\bf any EMRI detected by LISA with a circular, planar orbit {\em must} be caused by capture or migration within an AGN disc.} Prograde AGN-EMRI may display gas-drag effects early on in their evolution. We urge the LISA community to consider the AGN-EMRI channel when predicting EMRI waveforms and detection rates.

\subsection{TDE and AGN high states}\label{sec:Discussion:TDE}
From Fig.~\ref{fig:capturetimeST}, stars at most inclinations within $\mathcal{O} (10^{3-4} ~{\rm R_{g}})$ are captured by AGN discs over $\tau_{\rm AGN}$. Thus, we should expect a continuous supply of NSC stars to add to any initial embedded population. Retrograde stars end up captured at small disc radii and will experience scattering among the (typically more massive) sBH population growing there. A star that is scattered as a result of a close, chaotic dynamical encounter with a binary \citep{Wang21} or in a large-angle single-single scattering event, could be scattered into the AGN loss-cone, potentially yielding an AGN-TDE, which can be distinguished from ‘naked’ TDE by their lightcurves \citep{McKernan21}. Stars that are not tidally disrupted may experience Roche-lobe overflow (RLOF) onto the SMBH, yielding a temporary AGN high-state, possibly including quasi-periodicity \citep{Metzger22}. Stars may also be tidally disrupted by sBH in ‘micro-TDE’ \citep{Yang22}. Retrograde stars are far more efficiently captured by denser (SG-like) discs, so we expect RLOF events or TDE should be far more common in discs with gas densities $\geq 10^{-11}~{\rm g~cm^{-3}}$ --which may correspond to the more luminous Seyferts and quasars. Among non-AGN galaxies, TDE may preferentially occur in E+A galaxies \citep{French16}. If E+A galaxies are post-AGN galaxies then the excess of TDE could correspond to dynamical scatterings during relaxation among the post-AGN nuclear population, including the population discussed here. 

Embedded retrograde stars will experience orbital decay (‘starfall’) and end up either as a TDE/RLOF event as above, or swallowed whole by the SMBH. Prograde stars that spend time in the AGN disc will accrete from the gas disc and may grow to become super-massive ($100-200$~\msun) \citep{Cantiello20,Dittmann21}. Such stars must migrate within the disc and encounter other embedded objects, possibly merging with them \citep{Jermyn22}. If such stars are no longer surrounded by AGN disc gas, e.g. if they end up stalled in the innermost disc, they can lose mass rapidly and drive volume-filling outflows, which may contribute to the broad-line winds and outflows observed in AGN \citep{Jermyn22}. These ‘immortal’ prograde stars will have a significant impact on the disc locally and should be included in realistic disc models.

\subsection{Red Giants}\label{sec:Discussion:RG} 
Since Red Giants (RG) are geometrically large, we should expect that drag due to disc-crossing orbits is efficient. However, since RGs are diffuse and loosely bound, we should also expect that RGs lose much of their outer envelope during disc-crossing, especially in dense AGN discs. Indeed a single RG can lose up to 10$\%$ of its mass for every disc crossing \citep{Kieffer16}. Thus we should expect that RGs do not survive the capture process. RGs may evolve to become more diffuse and simply add their total mass to the accretion disc, or they may survive as a stripped core in a process that is $\ll$~Myr. This AGN-stripping of RGs may explain the lack of Red Giants in our own galactic nucleus \citep{Zaya20b, Zaya20a}.

\section{Conclusions}\label{sec:Conclusions}
AGN discs (particularly dense discs) can be efficient at capturing stars and stellar-origin black holes (sBH) on disc-crossing orbits over their lifetimes. Here we show that nuclear cluster stars on inclined retrograde orbits $(i>90^{\circ})$ are driven towards prograde disc capture $(i \to 0^{\circ})$, experience a large drop in semi-major axis $(a)$ and are captured significantly faster than inclined prograde stars. This population of captured stars may prompt detectable changes in AGN including AGN-TDE (around SMBH), AGN high-states due to Roche lobe overflow and micro-TDE (around sBH). Dense AGN discs are more efficient at stellar capture than less dense discs, as expected.

We find a critical angle $i_{\rm ret} \sim 113^{\circ}$, below which retrograde sBH decay towards fully embedded prograde orbits $(i\to 0^{\circ})$, while for $i_{\rm o}>i_{\rm ret}$ sBH decay towards fully embedded retrograde orbits $(i\to 180^{\circ})$. It is unclear whether $i_{\rm ret}$ is a universal critical angle, since some physics may be missing in our treatment (e.g. we consider exclusively circular orbits and we neglect the disc back-reaction). We shall investigate this phenomenon in future work. Prograde sBH experience the largest decrease in $a$ during disc capture. Larger AGN discs should capture a large number of sBH, with presumably a larger associated sBH-sBH merger rate and EMRI rate. More massive sBH are captured fastest by AGN discs, implying that larger mass sBH spend more time in AGN discs. As a result, larger mass sBH should have a greater bias towards spin alignment with the gas disc due to accretion gas torques. 

\section*{Acknowledgements}\label{sec:Acknowledgements}
SSN is supported by NSF PHY-2011334, and thanks Mordecai-Mark Mac Low for computational resources. GF is supported by ERC Starting Grant No.~\#~1218171001 and thanks Johan Samsing for insightful discussions. A.S. is supported by the National Science Foundation Graduate Research Fellowship Program under grant No.~\#~DGE1656466. KESF \& BM are supported by NSF AST-1831415 and Simons Foundation Grant 533845. JMB is supported by NSF award AST-2107764. NWCL gratefully acknowledges the generous support of a Fondecyt Iniciaci\'on grant 11180005, as well as support from Millenium Nucleus NCN19-058 (TITANs) and funding via the BASAL Centro de Excelencia en Astrofisica y Tecnologias Afines (CATA) grant PFB-06/2007. He also thanks support from ANID BASAL projects ACE210002 and FB210003.

\section*{Data Availability}
The data underlying this article will be shared on request to the corresponding author.

\bibliographystyle{mnras}\bibliography{refs}
\bsp\label{lastpage}\end{document}